Comment on "How perception of learning environment predicts male and female students' grades and motivational outcomes in algebra-based introductory physics courses "

M. B. Weissman

*Department of Physics, University of Illinois at Urbana-Champaign*

*1110 West Green Street, Urbana, IL 61801-3080*

Abstract

A recent paper used structural equation modeling to infer effects of gender-dependent student attitudes on several outcomes of introductory physics courses. The model used is precisely Markov equivalent to an equally plausible approximation in which a key gender-dependent coefficient in determining attitudes is indistinguishable from zero. I also argue that the qualitative framing may attribute to students motivations that actually belong to their teachers.

In order to improve education one needs to choose outcomes that are worth changing and to determine which causal factors both have a major effect on those outcomes and are susceptible to change by changing educational practices. Identifying important causal pathways is needed to pick which changes have a good chance to work. Should one seek interventions that will focus on interest on physics, or self-confidence in working problems, or sense of being comfortable in a group, or other attitudes or skills? Physics education researchers have not always been careful about how to identify such causal paths.(*1, 2*)  In this Comment I shall illustrate this issue by discussing a recent paper(*3*) in which a causal effect was inferred although the data presented were consistent with the effect being negligible. The same paper also made a common but questionable implicit assumption about which goals were important for the students in the study.

Cwik and Singh (*3*) examine gender differences in various outcomes in an introductory two-semester algebra-based physics course intended mainly for bioscience majors, the sort of course that pre-medical students typically take. In particular they claim that several forms of "students' perceptions of their learning environment" play an important role in determining not only "motivational outcomes" but also grades.

Gender differences in outcomes are directly measured and are described in Table III of the paper (*3*) without the need for causal models. The difference between the grades obtained by males and females ended up being small, with Cohen d in the second semester of 0.12, below the conventional statistical significance cutoff. (The much larger value given for the first semester appears to be a typographical error.) Attitudinal differences remained large, especially for self-reported interest, for which Cohen d was 0.71 in the first semester and 0.73 in the second.

Before getting into the more technical aspects of the analysis of how the gender differences arise, a brief remark on the attitudinal goals may be useful, since deciding which goals are worth reaching is as important as discovering what interventions would be most effective at reaching those goals. Cwik and Singh(*3*) assume that a key goal is for each student to acquire a physics *Identity*, i.e. to see themself as a "physics person".  This Identity outcome, unlike grades, is significantly different for males and females, with Cohen d of  0.45. The paper says "Studies have shown that it can be more difficult for women to form a physics identity than men." (*3*) Thus the paper treats the rather low rate at which these biology-oriented students, particularly females, identify as "physics persons" as resulting from the difficulty they face in achieving that identity. No evidence is given that the students themselves are trying to identify with physics. As members of the physics profession we have reasons to hope that more students identify with physics, but our difficulty in reaching that goal should not be confused with students' difficulties in reaching their own goals.

As is common in physics education research, Cwik and Singh (*3*) use linear structural equation models (SEMs) (*4*) to describe relations among several variables including gender, incoming SAT math scores, high school GPA, course grades, and attitudes measured by surveys. Although at most points the paper avoids explicitly causal language(*5*), the causal interpretation of the relations observed is explicit at key points. E.g. the abstract concludes "…perception of the learning environment predicts the outcomes and can be useful for creating an equitable and inclusive learning environment to help all students excel in these algebra-based physics courses." Only a cause, not a mere predictor, can "help" in "creating" good outcomes.

The simplest of the models used, their Fig. 2, is reproduced here as Fig. 1. Single-headed arrows in the Acylic Directed Mixed Graph (ADMG) (*6*) used to represent the model represent causal effects from tail to head variables. Double-headed arrows represent causal effects from unmeasured latent variables.

The key question these SEM models attempt to address is by what pathways gender leads to different outcomes. For example, does the appreciable gender difference in perceived recognition arise mainly from traits that are present before starting the course or from the "learning environment" of the course? Which traits are most important? Answers to such questions do not directly tell us at which stages interventions to alter outcomes might be most effective, but they can suggest which interventions might be good ones to investigate.

It is likely that no SEM representation of the sort used will come close to describing the actual causal relations among traits that develop over time with each variable potentially influencing subsequent values of the others. Nevertheless, for the sake of argument here we can just look at the analysis used within the SEM framework used by Cwik and Singh and in many other PER papers. At the simplest technical level, it is important to realize that the ADMG representation chosen is not determined by the measured correlations but requires some subjective choices about which arrows to use. Other ADMG representations can fit exactly equally well, i.e. are Markov equivalent (*6*), or approximately as well.

For those unfamiliar with reading SEMs, it may be useful to give a rough idea of the importance of the effects shown. In the model of Fig. 1, the effect of gender on the subjective sense of *Self efficacy* accounts for $(0.22*0.14)^2 = 0.001$ of the variance of *Perceived Recognition*. A larger gender effect flows through *Interest*: 0.027 of that variance.

The model of Fig. 1 shows no direct effects of *Gender* on any variables downstream from the second layer, i.e. after the first six variables on the left side. That means that after any possible effects on first-semester *Interest* and *Self efficacy 1* gender differences in the course learning environment do not show up here as significant causes. It is unlikely that much of the *Interest* and *Self efficacy 1* differences developed during the course since large differences are known to exist earlier in school(*7*) and prior work from the Singh group showed almost as big gender differences as these at the start of the first semester in a cohort selected to be more interested in physical science.(*8*) Thus the data presented do not imply that the college learning environment is adding to gender imbalances already predictable from prior traits. We shall see that the data also do not tell us the relative importance of *Interest* and *Self efficacy*.

Without attempting a full analysis of the many-variable correlation matrix (which is not quite given in complete form(*3*)) we can illustrate the sorts of alternatives available by exploring just the first six variables, the part shared by all four graphical models shown in the paper and which contains all the direct effects of *Gender*.

Since *Interest* remains highly stable over the two semesters and has by far the largest normalized gender difference of all the variables, it seems reasonable to see whether it could be the main attitude driving other gender differences in attitudes. It is quite plausible that a student's sense of whether they wish to work through problems on their own (*Self efficacy 1*) will be in large part a result of whether they are interested in the topic. The corresponding Markov equivalent ADMG would replace the double-headed latent variable connecting *Interest* and *Self efficacy 1* by a directed arrow from *Interest* to *Self efficacy 1*.

Although a Comment cannot replace primers or texts(*4*) on techniques, perhaps a brief description of how to calculate the coefficients in an alternative ADMG would be helpful. Measurement gives correlation coefficients among the three variables *Gender* (G), *Interest* (I), and *Self efficacy 1* (S): r(G,I), r(G,S), and r(I,S), i.e. dot products of the normalized vectors representing these variables. How these are expressed in terms of causal coefficients, e.g. c(G,I), depends on the ADMG chosen for the SEM. In the SEM of Fig. 1, we have r(G,I)=c(G,I) because there is only one path from G to I. There are two paths from G to S, one direct and one through math SAT (M), so r(G,S)= c(G,S)+c(G,M)*c(M,S).

The reason that our new ADMG is equivalent to the original one is that it still has three adjustable coefficients to fit three correlations. In the new SEM (with its causal coefficients denoted c') another directed path is added so

r(G,S)= c'(G,S)+c(G,M)*c(M,S)+c(G,I)*c'(I,S).

All causal terms here are just the same as in the initial SEM except for c'(G,S) and c'(I,S). We now have c'(G,S)=c(G,S)-c(G,I)*c'(I,S). The term c'(I,S) leads to the residual part of r(I,S) that is not accounted for by their shared ancestor G, i.e. that is from components of the vectors that are orthogonal to G. This residual is (r(I,S)-c(G,I)*(c(G,S)+c(G,M)*c(M,S))). The contribution to this residual comes only from the component of I that is orthogonal to G, whose squared magnitude is $(1-r(G,I)^2)$, so it is $c'(I,S)*(1-r(G,I)^2)$. Thus

$c'(I,S)*(1-r(G,I)^2)$= (r(I,S)-c(G,I)*(c(G,S)+c(G,M)*c(M,S))). Then in terms of the variables given in Fig.1:

c'(G,S)= c(G,S)-c(G,I)*(r(I,S)-c(G,I)*(c(G,S)+c(G,M)*c(M,S)))/$(1-c(G,I)^2)$=0.01.

The value 0.01 for the direct causal coefficient from *Gender* to *Self efficacy 1* is not only negligible but also well below the conventional 95% confidence statistical error bars (roughly ±0.07) and thus would be dropped altogether from the model as insignificant, following the standard procedure used by Cwik and Singh. The result would be a more parsimonious model that would not be Markov equivalent but rather better than the original because it has one less parameter and almost exactly as good a fit.

One might ask whether replacing the double-headed arrow between *Interest* and *Self efficacy 1* by a directional one from *Self efficacy 1* to *Interest* would be equally plausible. That would not be Markov equivalent to the initial ADMG because *SAT Math* is an ancestor of *Self efficacy 1* but not of *Interest*. Since *Gender* is much more strongly correlated with *Interest* than with *Self efficacy 1* it would not allow the arrow from *Gender* to *Interest* to be removed. It would probably require adding an arrow with a significant negative coefficient from *SAT Math* to *Interest*, making a less parsimonious model with an intuitively implausible negative effect.

Nevertheless other graphs can be chosen even for these first few variables. For example, Fig. 1 has a two-headed arrow between *Physics 1 Grade* and *Self efficacy 1,* representing an assumption that some unmeasured latent variable affects both and that how well a student is doing in a course does not affect their sense of self efficacy, and vice versa. That arrow could be replaced by a single-headed arrow from the former to the latter. The meaning of that arrow would be that students who are doing well in the course tend to develop enhanced confidence. The result would not be Markov equivalent to the model chosen, but would appear to fit the data about as well. No doubt a variety of similar choices are also possible for the later stages of the graph. (Here it is important to remember that it is not realistic to represent traits that develop over time while influencing each other as if they were single-time variables in an ADMG.)

My point is that the coefficients given are highly dependent on what model is chosen for the underlying causal pattern. Conclusions about which traits are causally important flow not from the data but from assumptions implicit in the graphical pattern chosen. That choice is provably not dictated by the data. Although some model choices are influenced by subjective interpretations of previous interviews with small numbers of students (e.g. (*9*) ) the correlational data do not add new evidence to support those choices.

The choice of which variables to include can express important assumptions even before causal assumptions become imbedded in a choice of ADMG connections. Omitting a variable that mediates gender effects can lead to spurious coefficients for direct gender effects on downstream results regardless of whether including that variable significantly improves the overall predictive accuracy. Omitting *SAT Math, High School GPA, Interest,* or *Self Efficacy 1* would require that other variables pick up direct effects from Gender to replace the effects mediated by the

2/17/23 12:06 PM                                                                                                                              6

earlier variables. For example, dropping *Self Efficacy 1* from the model shown in Fig. 4 of Cwik and Singh would require adding some direct effect from *Gender* to *Belonging*, since *Self Efficacy 1* mediates most of the causal flow from *Gender* to *Belonging* in that SEM. Thus dropping early mediating variables can create the impression that gender effects on later variables are mediated by subsequent college learning environment more than would be found in a more complete analysis.

Recent follow-up papers(*10, 11*) look at similar attitudinal results but omit the early variables *Interest* and *Self Efficacy*. *Interest* is easily the most gender-asymmetric of the traits described in (*3*) and *Self Efficacy 1* is second largest. Thus dropping them can have particularly large consequences for coefficients of direct downstream effects of *Gender*. This matters because college "learning environment" is suspected of playing a mediating role in creating the gender differences for which such earlier mediators are not found. In other words, the case for attributing gender differences in outcomes to ways that college teaching unequally affects males and females rests on the extent to which the different outcomes are not predictable from pre-college differences (measured by math SAT and high school GPA) or from differences in interest that may have existed before college(*7*). It is possible that if interest were not excluded from the analysis the newer results would also show little evidence of a gender effect via routes other than interest, grades and math SATs, as in this study(*3*).

Unfortunately the results of these studies that omit the *Interest* variable have been reported in the scientific press(*12*) with policy recommendations based on a purely causal interpretation. The press report even omits key math SAT score data that had been in in the original papers(*3, 10, 11*). That press report was then featured on the Physical Review Physics Education Research front page. The message is that mid-stream attitudinal variables are crucial, but we have seen that the data themselves do not tell us whether interventions focused on them would have any noticeable effect on the correlation between gender and downstream outcomes.

The case made by Cwik and Singh(*3*) for focusing effort on mid-stream attitudinal variables thus has at least two major weaknesses. One is the lack of justification for an implicit assumption about goals- that these students want to identify as "physics persons" or at least ought to want that. The other is the lack of support in the data for claims about causal pathways to the desired outcomes. Fortunately, at least one experimental study has found benefits of some gender-motivated classroom changes.(*13*) More such experimental studies would be worthwhile.

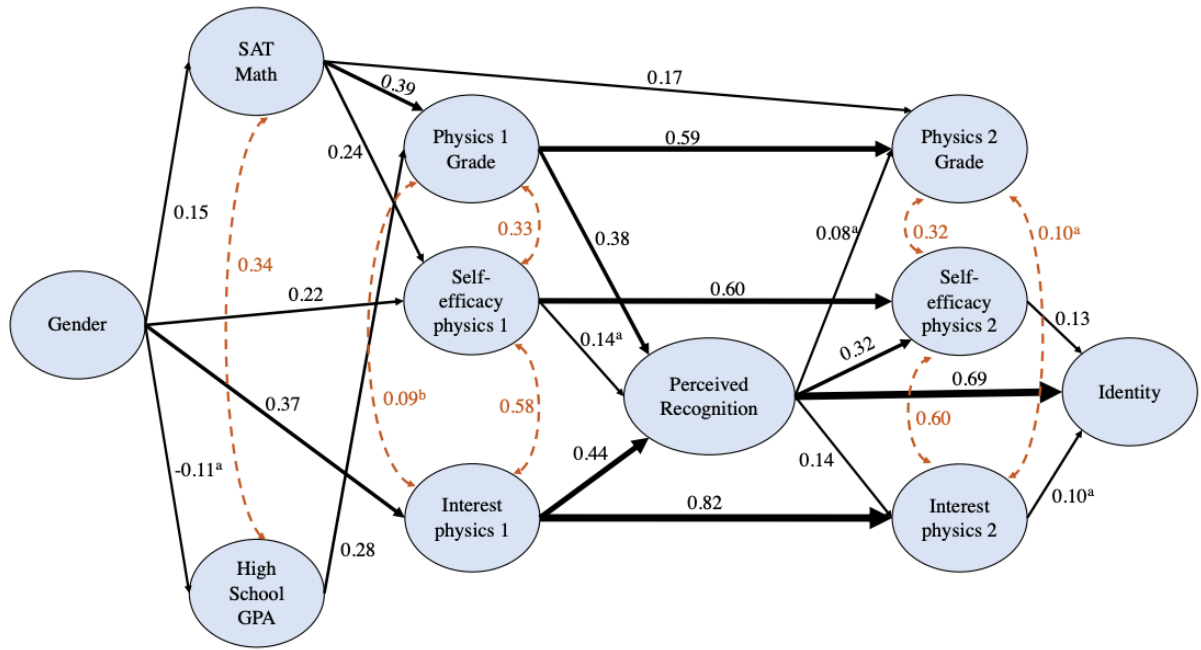

Fig. 1. This is a reproduction of Fig. 2 of (*3*). The coefficients given for the single-headed arrows are regression coefficients for dependences of the normalized arrowhead variables on the normalized arrow-tail parent variables, holding other parents of the arrowhead variable constant. The coefficients given for the double-headed dashed arrows are simple correlation coefficients. An implicit residual coefficient due to the effects of a latent variable on each of the measured pair can be calculated by subtracting the correlation due to the directed arrows. (A more consistent and transparent notation would instead just give the residual correlation coefficient, i.e. the component of correlation beyond that already implied by the upstream part of the graph.)